\documentclass[longauth]{aa}  

\usepackage{graphicx}
\usepackage{xcolor}
\graphicspath{{./}}

\usepackage{amsmath}
\usepackage{enumitem}
\usepackage{txfonts}
\usepackage{multirow}
\usepackage[hyperfootnotes=false]{hyperref}
\hypersetup{colorlinks=True,linkcolor=black,citecolor=blue}
\newcommand{\cii}{[C\,{\sc ii}]}

\newcommand{\hi}{H\,{\sc i}}

\newcommand{\ciil}{[C\,{\sc ii}] 158\,$\mu{\rm m}$}

\begin{document}

   \title{How much gas and dust is in the $z=5.7$ Lyman Break Galaxy HZ10?
   }
   \subtitle{An ALMA Band 10 to 4 and JWST/NIRSpec study of its interstellar medium
   }

   \author{H. S. B. Algera
          \inst{1}$^{\star}$,
          R. Herrera-Camus
          \inst{2,3}\thanks{Both authors contributed equally to this work.}
          \and 
          M. Aravena
          \inst{4,3}
          \and 
          R. Assef
          \inst{4}
          \and
          T. L. J. C. Bakx
          \inst{5}
          \and 
          A. Bolatto
          \inst{6}
          \and
          K. Cescon
          \inst{7}
          \and 
          C.-C. Chen
          \inst{1}
          \and 
          E. da Cunha
          \inst{8}
          \and
          P. Dayal
          \inst{9,10,11}
          \and 
          I. De Looze
          \inst{12}
          \and 
          T. Diaz-Santos
          \inst{13,14}
          \and         
          A. Faisst
          \inst{15}
          \and
          A. Ferrara
          \inst{16}
          \and 
          N. Förster Schreiber
          \inst{17}
          \and 
          N. Hathi
          \inst{18}
          \and 
          R. Ikeda
          \inst{19,20}
          \and 
          H. Inami
          \inst{21}
          \and
          G. C. Jones
          \inst{22,23}
          \and 
          A. Koekemoer
          \inst{18}
          \and 
          D. Lutz
          \inst{24}
          \and 
          M. Relaño
          \inst{25,26}
          \and 
          M. Romano
          \inst{27,28}
          \and 
          L. Rowland 
          \inst{7}
          \and 
          L. Sommovigo
          \inst{29}
          \and 
          L.\ Vallini
          \inst{30}
          \and 
          A. Vijayan
          \inst{31}
          \and
          V. Villanueva
          \inst{2}
          \and 
          P. van der Werf 
          \inst{7}
          }

   \institute{Institute of Astronomy and Astrophysics, Academia Sinica, 11F of Astronomy-Mathematics Building, No.1, Sec. 4, Roosevelt Rd, Taipei 106319, Taiwan, R.O.C.\\ 
              \email{hsbalgera@asiaa.sinica.edu.tw}
         \and
             Universidad de Concepción, Víctor Lamas 1290, Barrio Universitario, Concepción, Chile
        \and
             Millennium Institute for Galaxies (MINGAL)
        \and
            Instituto de Estudios Astrof\'{\i}cos, Facultad de Ingenier\'{\i}a y Ciencias, Universidad Diego Portales, Av. Ej\'ercito 441, Santiago, Chile
        \and
            Department of Space, Earth, \& Environment, Chalmers University of Technology, Chalmersplatsen, SE-4 412 96 Gothenburg, Sweden
        \and
            Department of Astronomy and Joint Space-Science Institute, University of Maryland, College Park, MD 20854
        \and
            Leiden Observatory, Leiden University, NL-2300 RA Leiden, the Netherlands
        \and
            International Centre for Radio Astronomy Research, University of Western Australia, 35 Stirling Hwy, Crawley 26WA 6009, Australia
        \and 
            Canadian Institute for Theoretical Astrophysics, 60 St George St, University of Toronto, Toronto, ON M5S 3H8, Canada
        \and
            David A. Dunlap Department of Astronomy and Astrophysics, University of Toronto, 50 St George St, Toronto ON M5S 3H4, Canada
        \and 
            Department of Physics, 60 St George St, University of Toronto, Toronto, ON M5S 3H8, Canada
        \and 
            Sterrenkundig Observatorium, Ghent University, Krĳgslaan 281 S9, B-9000 Ghent, Belgium
        \and 
            Institute of Astrophysics, Foundation for Research and Technology Hellas (FORTH), Heraklion, GR-70013, Greece
        \and 
            School of Sciences, European University Cyprus, Diogenes street, Engomi, 1516 Nicosia, Cyprus
        \and 
            Caltech/IPAC, MS 314-6, 1200 E. California Blvd. Pasadena, CA 91125, USA
        \and 
            Scuola Normale Superiore, Piazza dei Cavalieri 7, 56126, Pisa, Italy
        \and 
            Max-Planck-Institut für Extraterrestiche Physik (MPE), Giessenbachstr., 85748, Garching, Germany
        \and 
            Space Telescope Science Institute, 3700 San Martin Drive, Baltimore, MD 21218, USA
        \and 
            Department of Astronomical Science, SOKENDAI (The Graduate University for Advanced Studies), Mitaka, Tokyo 181-8588, Japan
        \and
            National Astronomical Observatory of Japan, 2-21-1 Osawa, Mitaka, Tokyo 181-8588, Japan
        \and
            Hiroshima Astrophysical Science Center, Hiroshima University, 1-3-1 Kagamiyama, Higashi-Hiroshima, Hiroshima 739-8526, Japan
        \and
            Kavli Institute for Cosmology, University of Cambridge, Madingley Road, Cambridge CB3 0HA, UK
        \and 
            Cavendish Laboratory, University of Cambridge, 19 JJ Thomson Avenue, Cambridge CB3 0HE, UK
        \and
            National Radio Astronomy Observatory, 520 Edgemont Road, Charlottesville, VA 22903, USA
        \and 
            Dept. Fisica Teorica y del Cosmos, Universidad de Granada, Granada, Spain
        \and 
            Instituto Universitario Carlos I de Física Teórica y Computacional, Universidad de Granada, E-18071 Granada, Spain
        \and
            Max-Planck-Institut f\"{u}r Radioastronomie, Auf dem H\"{u}gel 69, 53121 Bonn, Germany
        \and 
            INAF – Osservatorio Astronomico di Padova, Vicolo dell’Osservatorio 5, I-35122 Padova, Italy
        \and 
            Center for Computational Astrophysics, Flatiron Institute, 162 5th Avenue, New York, NY 10010, USA
        \and
            INAF – Osservatorio di Astrofisica e Scienza dello Spazio di Bologna, Via Piero Gobetti, 93/3, I-40129 Bologna, Italy
        \and 
            Astronomy Centre, University of Sussex, Falmer, Brighton BN1 9QH, UK
             }

    \date{Received -; accepted -}

  \abstract
  {A complete overview of the stellar, gas and dust contents of galaxies is key to understanding their assembly at early times. However, an estimation of molecular and atomic gas reservoirs at high redshift relies on various indirect tracers, while robust dust mass measurements require multi-band far-infrared continuum observations.}
  {We take census of the full baryonic content of the main-sequence star-forming galaxy HZ10 at $z=5.65$, a unique case study where all necessary tracers are available.}
  {We present new ALMA Band 10 ($\lambda_{\rm rest}\approx50\,\mu\mathrm{m}$) and Band 4 ($\approx300\,\mu\mathrm{m}$) observations towards HZ10, which combined with previously taken ALMA Band 6 through 9 data ($\approx70 - 200\,\mu\mathrm{m}$) constrains its dust properties. We complete the baryonic picture using archival high-resolution \cii{} observations that provide both a dynamical mass and molecular and atomic gas mass estimates, a JVLA CO(2-1)-based molecular gas mass, and \textit{JWST} metallicity and stellar mass measurements.} 
  {We detect continuum emission from HZ10 in Bands 10 and 4 at the $3.4-4.0\sigma$ level, and measure a dust temperature of $T_\mathrm{dust} = 37_{-5}^{+6}\,$K and dust mass $\log(M_\mathrm{dust}/M_\odot) = 8.0 \pm 0.1$. Leveraging the dynamical constraints, we infer its total gas budget, and find that commonly used \cii{}-to-H$_2$ and \cii{}-to-HI conversions overpredict the gas mass relative to the dynamical mass. For this reason, we derive a \cii{}–to–total ISM mass (atomic + molecular) conversion factor, which for HZ10 corresponds to $\alpha_{\rm [CII]}^{\rm ISM} = 39^{+50}_{-25}~M_{\odot}~L_{\odot}^{-1}$. We also find that HZ10 falls below the local scaling relation between dust-to-gas ratio and metallicity, suggesting inefficient ISM dust growth.} 
  {These results demonstrate a powerful synergy between ALMA and \textit{JWST} in disentangling the baryonic components of early galaxies, paving the way for future studies of larger samples to dynamically calibrate \cii{}-to-gas mass conversion factors, and further unravel the pathways of early dust build-up.}


   \maketitle
%
%-------------------------------------------------------------------

\section{Introduction}
\label{sec:introduction}

The evolution of galaxies is driven by both dark matter, which establishes the initial gravitational potential that accretes primordial gas, and baryonic matter, whose complex interplay leads to star formation. A complete census of the baryonic components—gas, dust, and stars—is therefore essential to understanding galaxy evolution across cosmic time.

The atomic and molecular gas components play a central role in this process, serving as the raw material for star formation \cite[e.g., ][]{kennicutt12,krumholz12}. The atomic gas reservoir can be directly traced by the 21~cm emission line of neutral hydrogen (\(\rm H I\)), but this line becomes inaccessible due to instrumental limitations already at $z\gtrsim0.5 - 1$ (e.g., \citealt{fernandez2016}). Molecular hydrogen, the dominant form of molecular gas and the direct fuel for star formation, is commonly traced by low-\(J\) rotational transitions of carbon monoxide (CO). However, CO-based gas mass estimates depend on the CO-to-\(\rm H_2\) conversion factor (\(\alpha_{\rm CO}\)), which is known to vary with metallicity and gas density \cite[e.g.,][]{bolatto2013,chiang2024,teng2024}.  In low-metallicity environments, CO emission may be weak or absent due to photodissociation, leaving a significant fraction of molecular gas undetectable in CO-based studies, commonly referred to as ``CO-faint'' or ``CO-dark'' molecular gas \cite[e.g.,][]{wolfire10,madden2020}. At high redshift, CO is furthermore difficult to detect against the warm Cosmic Microwave Background (CMB; \citealt{dacunha2013,frias-castillo2024}).

The \ciil\ fine-structure line is expected to trace both atomic and molecular gas in galaxies, with a minor fraction of \cii{} emission moreover thought to arise from the ionized interstellar medium \citep[ISM; e.g.,][]{croxall2017,diaz-santos2017,zhao2024}. Given its typical brightness and multi-phase origin, \cii{} is a particularly valuable tool for studying the ISM of high-redshift systems, where direct observations of CO and \hi\ are often challenging. For instance, \cite{heintz21} used \hi\ absorption in $\gamma$-ray burst host galaxies at $z \gtrsim 2$ to directly measure the \cii-to-\hi\ conversion factor (\(\beta_{\rm HI}\)). For molecular gas, various methods have been employed to derive a \cii-to-H$_{2}$ conversion factor in both low-metallicity galaxies \citep[e.g.,][]{accurso17,jameson18,madden2020,hunt23} and high-redshift systems \citep[e.g.,][]{zanella18,pavesi2019,dessaugeszavadsky2020}. At the same time, various theoretical works have tried to calibrate \cii{} as a tracer of both H$_{2}$ and \hi\ in simulations \citep{vizgan2022a,vizgan2022b,gurman2024,casavecchia2025,vallini25}. Observational studies generally find that galaxies at $z\gtrsim4$ are rich in both atomic and molecular gas \cite[e.g.,][]{dessaugeszavadsky2020,heintz21,aravena2024}, similar to those at cosmic noon \cite[e.g., ][]{scoville17,tacconi2018,tacconi20}. However, discrepancies between different calibrations and methodologies remain a major source of uncertainty, and until recently, no simultaneous calibrations existed for atomic and molecular gas in high-redshift galaxies \citep[c.f., the theoretical works by][]{casavecchia2025,vallini25}. In this context, detailed observations of high-redshift systems—where baryonic mass can be inferred from kinematics alongside independent measurements of gas mass, stellar mass, and metallicity—are crucial for improving the accuracy of these estimates.

Dust—despite being negligible in terms of the total mass budget—is another important baryonic component in galaxies due to its significant effect on their observable spectral energy distribution (SED), and importance in catalyzing the formation of molecular hydrogen \citep[e.g.,][]{galliano2018}. Recent observations with the Atacama Large Millimeter/submillimeter Array (ALMA) have revealed the widespread presence of dust in the early Universe \citep[$z\gtrsim6$; e.g.,][]{tamura2019,inami2022}. However, to understand the exact pathways of early dust production, it is essential to precisely constrain the dust masses of high-redshift galaxies through multi-band ALMA continuum observations. In particular, observations probing both the peak and Rayleigh-Jeans tail of the dust SED are crucial to break degeneracies between the dust mass, temperature and emissivity index of high-redshift galaxies \citep[e.g.,][]{bakx2021,dacunha2021,algera2024b}. Covering the peak of the SED requires dedicated high-frequency observations, for which ALMA Band 9 -- tracing rest-frame $\lambda \lesssim 80\,\mu\mathrm{m}$ at $z\gtrsim 5$ -- has recently been employed to great effect \citep[e.g.,][]{bakx2021,bakx2025,algera2024b,tripodi2024,villanueva2024}. However, to obtain the most stringent constraints it is necessary to sample beyond the peak, requiring observations in the under-exploited ALMA Band 10 \citep[e.g.,][]{fernandez-aranda2025}.

Further insight into dust production mechanisms requires knowledge of galaxy metallicities, as metallicity appears to drive the dust-to-gas and dust-to-metal ratios of galaxies in the nearby Universe \citep[e.g.,][]{fisher2014,remyruyer2014,devis2019,delooze2020,galliano2021}. In particular, at low metallicities dust is thought to predominantly be produced by stellar sources—supernovae (SNe) and AGB stars—while at higher metallicities dust growth in the dense ISM likely takes over \citep[e.g.,][]{inoue2011,asano2013,feldmann2015,popping2017,galliano2021,choban2024,sawant2025}. With the advent of the \textit{James Webb Space Telescope} (\textit{JWST}), the first metallicity measurements of dusty galaxies at $z\gtrsim6$ have now become available \citep[e.g.,][]{valentino2024,rowland2025} and enable testing these low-redshift scaling relations at an epoch where dust has had significantly less time to form. The high dust masses and metallicities of UV-luminous galaxies at $z\sim7$ suggest rapid dust build-up through a combination of SNe and ISM dust growth \citep{algera2025}, although galaxies at even higher redshifts show lower dust-to-stellar mass ratios suggesting limited growth in the ISM (\citealt{ciesla2024,algera2025,burgarella2025,ferrara2025,mitsuhashi2025}; see also \citealt{ferrara2016} for a theoretical perspective on why ISM growth may be inhibited at early times). Larger galaxy samples with well-constrained dust masses from multi-band ALMA observations are required to more accurately test theoretical models of early dust production.

In this paper, we present the case study of HZ10, one of the best examples---if not the best---for examining the total mass budget in a normal, star-forming galaxy at $z \sim 6$. This is partly because HZ10 is, to date, the only main-sequence galaxy at this cosmic epoch with a combination of: $(i)$ a direct detection of molecular gas via a low-$J$ CO transition \citep[CO(2-1);][]{pavesi2019}, $(ii)$ a robust dynamical mass estimate based on detailed kinematic modeling of its \cii{} line emission \citep{telikova2024}, $(iii)$ a multi-band dust SED \cite[this work; ][]{faisst2020,villanueva2024}, and $(iv)$ a metallicity measurement from \textit{JWST}/NIRSpec observations \citep{jones2024}.

Throughout this work, we assume a standard flat $\Lambda$CDM cosmology with $\Omega_m = 1 - \Omega_\Lambda = 0.30$ and $H_0 = 70\,\mathrm{km/s/Mpc}$. At the redshift of HZ10, $z=5.65$, 1 arcsec corresponds to $5.9\,\mathrm{kpc}$. We moreover assume a \citet{chabrier2003} initial mass function (IMF).

%--------------------------------------------------------------------

\begin{figure}
    \centering
    \includegraphics[width=1.0\linewidth]{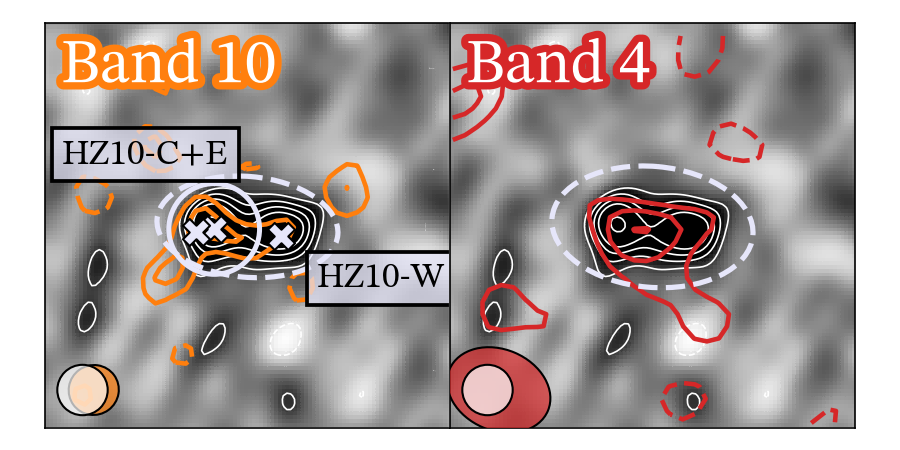}
    \caption{The new ALMA observations of HZ10 presented in this work. \textit{(Left)}: ALMA Band 10 contours (orange) on top of the ALMA Band 9 continuum imaging presented in \citet[][background greyscale and white contours]{villanueva2024}. The centroids of the central (C) and western (W) components as identified via \cii{} emission in \citet{telikova2024} are annotated. We moreover show the center of the eastern component (E) identified in \textit{JWST}/NIRSpec by \citet{jones2024}, although it is fully blended with HZ10-C at the resolution of the ALMA observations. \textit{(Right)}: ALMA Band 4 contours (red) on top of the ALMA Band 9 map. Both cutouts span $4''\times4''$, and beam sizes are shown in the lower left corner. Contours are placed at $\pm1\sigma$ intervals starting from $\pm2\sigma$ with negative contours being dashed. Band 10 (4) continuum emission is detected at the $3.4\sigma$ ($4.0\sigma$) level. We extract the Band 10 and 4 flux densities of the full HZ10 system in the dashed apertures, avoiding a likely noise spike to the south-east in the Band 10 map. We extract the Band 10 flux density of HZ10-C+E in the smaller circular aperture.}
    \label{fig:cutouts}
\end{figure}

\section{Target and data}
\label{sec:data}

\subsection{HZ10}
In this work, we present new ALMA Band 10 and 4 observations towards the massive \cite[$\log(M_{\star}/M_\odot)= 10.35\pm0.37$;][]{mitsuhashi2024} Lyman-break galaxy HZ10 in the COSMOS field \citep{scoville2007}. The redshift of HZ10 was initially measured spectroscopically via a detection of its Lyman-$\alpha$ emission ($z_{\mathrm{Ly}\alpha} = 5.659 \pm 0.004$) through Keck/DEIMOS observations, and the galaxy was subsequently followed up by ALMA in \ciil{} ($z_\text{\cii{}} = 5.6566 \pm 0.0002$) and underlying dust continuum emission in Band 7 \citep{capak2015}. Among the nine $z\sim5-6$ galaxies in the \citeauthor{capak2015} sample, HZ10 turned out to be the most FIR-luminous continuum source, as well as one of brightest \cii{} emitters. This motivated follow-up observations of its CO(2-1) line using the Very Large Array (VLA) by \citet{pavesi2019}, who detected the line at $>8\sigma$ significance. The CO(2-1) detection in HZ10 represents one of the most distant detections of molecular gas in a main-sequence galaxy to date (c.f., \citealt{dodorico2018,zavala2022}). 

To study its dust continuum emission, \citet{faisst2020} presented additional ALMA Band 6 and 8 observations towards HZ10. From a modified blackbody fit including the aforementioned Band 7 data, they determined its dust temperature and infrared luminosity to be $T_\mathrm{dust} = 46_{-8}^{+16}\,\mathrm{K}$ and $\log(L_\mathrm{IR}/L_\odot) = 12.5_{-0.3}^{+0.1}$, respectively, suggesting HZ10 to be an Ultra-luminous Infrared Galaxy (ULIRG; defined as having $L_\mathrm{IR} > 10^{12}\,L_\odot$). New ALMA Band 9 observations, which better capture the peak of the infrared emission, were recently presented by \citet{villanueva2024}. They refined the constraints on the dust SED of HZ10, finding a dust temperature of $T_\mathrm{dust} = 47 \pm 7\,\mathrm{K}$, consistent with \citet{faisst2020}.

Recent observations of HZ10 with \textit{JWST}/NIRSpec as part of the GA-NIFS program were presented by \citet{jones2024}. Their observations reveal a clumpy, three-component structure: a main component (HZ10-C) separated from a second minor component (HZ10-E) by only $\sim1.6$~kpc, and a third component (HZ10-W) separated by $\sim3.6$~kpc. \citet{telikova2024}, using high-resolution ($\theta\approx0.\arcsec3$) ALMA \cii\ observations, find evidence that the HZ10-C+E components correspond to a disturbed rotating disk with a rotational support of $V_{\rm rot}/\sigma_{0}\approx2$. The western component, HZ10-W, is both fainter in \cii{} emission and more compact, preventing a similarly detailed kinematic analysis. Finally, based on various emission line ratios, \citet{jones2024} find that HZ10 is primarily powered by star formation, and has a gas-phase metallicity of $Z\approx0.52 - 0.71\,Z_\odot$ across the three components (HZ10-C, -E and -W). Based on their findings, we adopt a metallicity of $Z/Z_\odot = 0.6 \pm 0.1$ in our analysis, both for HZ10 as a whole, and for the rotationally-supported HZ10-C+E system.

\subsection{New ALMA Band 4 and 10 observations}

\begin{figure*}[t]
    \centering
    \includegraphics[width=0.9\linewidth]{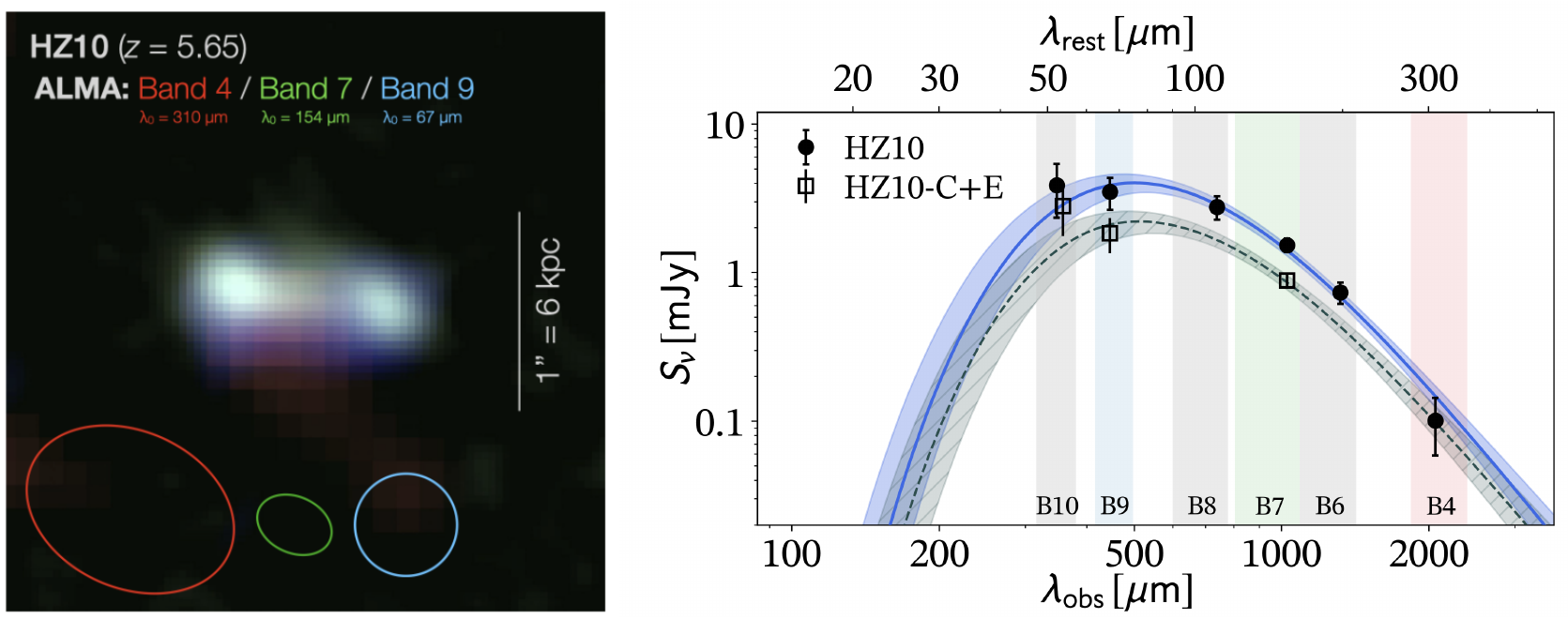}
    \caption{\textit{(Left)}: Composite image of HZ10 combining ALMA continuum observations from Band 4 (red), Band 7 (green), and Band 9 (blue). The beam sizes for each ALMA band are shown at the bottom, and the physical scale is indicated on the right. \textit{(Right)}: Our fiducial optically thin modified blackbody fits to the multi-band ALMA photometry of HZ10 (black circles and blue shading) and HZ10-C+E (open squares and dashed grey shading). The central wavelength of the Band 10 measurement for HZ10-C+E is shifted by $+10\,\mu\mathrm{m}$ for visual clarity. All ALMA bands used in this work are showing through the vertical shading, with those used in the composite image shaded in the same color.}
    \label{fig:dataFit}
\end{figure*}

We present new ALMA Band 10 ($0.35\,\mathrm{mm}$) and Band 4 ($2\,$mm) observations towards HZ10, taken as part of projects 2024.1.01134.S (PI Herrera-Camus) and 2023.1.01033.S (PI Algera), respectively. These new observations are shown in Figure \ref{fig:cutouts}, where they are overlaid on the ALMA Band 9 image presented in \citet{villanueva2024}.

The primary goal of the Band 4 observations was to probe the Rayleigh-Jeans tail of the dust emission and thereby improve the constraints on the dust mass and emissivity index of HZ10. Owing to its brightness, a short integration time of $t_\mathrm{int}\approx300\,$s was adopted in Band 4. The calibrated visibility data were restored in {\sc{casa}} version 6.5.4  \citep{mcmullin2007,casa_team2022} following the standard approach using {\sc{scriptForPI.py}}. We confirmed that the CO(8-7) line, redshifted to $\nu_\mathrm{obs}=138.47\,$GHz, was not detected, and therefore all channels were included in the imaging. We adopted natural weighting to maximize sensitivity, yielding a beam size of $1\farcs06\times0\farcs81$ and a root-mean-square (rms) sensitivity of $\sigma = 30.4\,\mu\mathrm{Jy\,beam}^{-1}$. Continuum emission from HZ10 is detected at the $\sim4.0\sigma$ level, with the peak of the emission being located in between the central (C+E) and western (W) components of the galaxy.

The primary goal of the ALMA Band 10 observations was to detect the [OIII]~52~$\mu$m transition, and the results will be presented in Herrera-Camus et al. (in preparation). The total Band 10 integration time was $t_\mathrm{int}=154.3$ min. The calibrated visibility data were restored in {\sc{casa}} version 6.6 following the standard approach using {\sc{scriptForPI.py}}. Using natural weighting, the Band 10 image attains a sensitivity of $\sigma = 536\,\mu\mathrm{Jy~beam^{-1}}$ with a beam size of $0\farcs38\times0\farcs35$. However, we opt to use a Band 10 image tapered to a circular beam of $0\farcs5$ to avoid over-resolving the emission of HZ10 and to match the resolution of the Band 9 data. We detect a $\sim3.4\sigma$ signal coincident with the position of HZ10-C+E, and a further $\sim2\sigma$ elongated feature stretching out to HZ10-W. Moreover, we observe a $\sim3\sigma$ emission feature stretching out from HZ10-C+E towards the south-east. As the significantly deeper Band 9 data do not show any signal at this location, we deem it to be a noise spike.

\subsection{Flux density measurements}
We extract the flux density of the entire HZ10 system in the elliptical apertures shown in Fig. \ref{fig:cutouts} (Bands 10 and 4) and Fig.\ \ref{fig:allCutouts} (all bands). The resulting measurements are provided in Table \ref{tab:properties} in Appendix \ref{app:properties}. We verify that the flux densities measured in Bands 6, 7, 8 and 9 are consistent with previous works \citep{faisst2020,villanueva2024}. The high angular resolution of the observations in Bands 7, 9 and 10 moreover enables a flux density measurement of just HZ10-C+E, the rotationally supported component of the HZ10 system. We extract these measurements in a smaller aperture centered on the centroid of the \cii{} emission of HZ10-C+E reported in \citet{telikova2024}.

To account for systematic uncertainties in the ALMA flux scale, we add additional calibration uncertainties in quadrature to the measurement errors, following the ALMA technical handbook.\footnote{\url{https://almascience.nao.ac.jp/documents-and-tools/cycle11/alma-technical-handbook}} For ALMA Band 4 (Bands 6, 7 and 8), we adopt a $5\%$ ($10\%$) calibration uncertainty, while for Bands 9 and 10 we adopt a larger recommended systematic error of $20\%$.

\section{Results}
\label{sec:results}

\begin{table}[t]
\centering
\caption{Optically thin modified blackbody fitting results for the integrated photometry and resolved C+E components of HZ10.}
%\resizebox{\linewidth}{!}{%
\begin{tabular}{l|cc}
\hline\hline
& \textbf{HZ10} & \textbf{HZ10-C+E} \\
\hline

$\log(M_\mathrm{dust}/M_\odot)$ & $8.0 \pm 0.1$ &  $7.9 \pm 0.2$ \\
$T_\mathrm{dust}\,\mathrm{[K]}$ & $37_{-5}^{+6}$ & $37_{-5}^{+6}$ \\
$\beta_\mathrm{IR}$ & $2.2 \pm 0.4$ & $2.1 \pm 0.4$\\
$\lambda_\mathrm{peak}/\mu\mathrm{m}$ & $76_{-6}^{+7}$ & $ 78_{-7}^{+8}$ \\
$\log(L_\mathrm{IR}/L_\odot)$ & $12.4 \pm 0.1$ & $12.1 \pm 0.1$ \\
\hline\hline 
\end{tabular}
\label{tab:mbb}
\end{table}

The left panel of Figure \ref{fig:dataFit} shows a composite image of HZ10 combining ALMA continuum observations from Bands 4, 7, and 9. In this image, we clearly distinguish the two main components, HZ10-C+E and HZ10-W. As shown by the beam size achieved in the different ALMA bands at the bottom of the figure, only the Band 7 and Band 9 observations spatially resolve HZ10 \citep{villanueva2024,telikova2024}. Thus, to incorporate the Band 4 observations into the dust SED modeling, we first focus on the source-integrated properties of HZ10. For completeness, we also present the images in all of Bands 10 - 4 in Figure \ref{fig:allCutouts} in Appendix \ref{app:cutouts}, and provide the flux measurements as well as other relevant physical parameters in Table \ref{tab:properties} in Appendix \ref{app:properties}.

We fit the source-integrated continuum photometry of HZ10 with a single-temperature, optically thin modified blackbody (MBB; right panel of Fig.\ \ref{fig:dataFit}), following the fitting framework introduced in \citet{algera2024a}. We leave the dust mass, temperature and emissivity index as free parameters and adopt flat priors on $\log(M_\mathrm{dust}/M_\odot)$ and $T_\mathrm{dust}$ following \citet{algera2024a,algera2024b}. Given the relatively low S/N of the Band 4 detection, we adopt a Gaussian prior on $\beta_\mathrm{IR}$ with a mean (standard deviation) of $2.0$ ($0.5$). This range is consistent with recent determinations of $\beta_\mathrm{IR}$ in submillimeter galaxies (e.g., \citealt{dacunha2021,liao2024,bendo2025}), as well as in galaxies and quasars at $z\gtrsim4$ \citep{witstok2023,algera2024b,tripodi2024}. A dust mass opacity coefficient $\kappa_\nu$ at rest-frame $850\,\mu\mathrm{m}$ of $\kappa_{850} = 0.77\,\mathrm{g}^{-1}\,\mathrm{cm}^{2}$ is assumed, following \citet{dunne2000}. The effect of the CMB is moreover included in the fit, following \citet{dacunha2013}, and dust temperatures and IR luminosities are quoted as if at $z=0$ (i.e., subject to a CMB temperature of $2.73\,\mathrm{K}$).

We adopt the median and $16-84^\mathrm{th}$ percentiles as the `best fit' and corresponding uncertainty, respectively, and present the results in Table \ref{tab:mbb}. From the fit, we infer a dust temperature of $T_\mathrm{dust} = 37_{-5}^{+6}\,$K, a dust mass of $\log(M_\mathrm{dust} / M_\odot) = 8.0 \pm 0.1$, and a dust emissivity index of $\beta_\mathrm{IR} = 2.2 \pm 0.4$.\footnote{We note that, had we assumed a flat prior on $\beta_\mathrm{IR}$ between $1 < \beta_\mathrm{IR} < 4$ following \citet{algera2024b}, we would recover a slightly steeper but overall consistent value of $\beta_\mathrm{IR}= 2.5_{-0.5}^{+0.6}$.} The corresponding dust SED peaks at a wavelength of $\lambda_\mathrm{rest} = 76_{-6}^{+7}\,\mu\mathrm{m}$, i.e., between ALMA Bands 8 and 9. Integrating the best-fit MBB across rest-frame $8-1000\,\mu\mathrm{m}$ yields a total infrared luminosity of $\log(L_\mathrm{IR}/L_\odot) = 12.4\pm0.1$, which classifies HZ10 as a ULIRG, consistent with previous works \citep[e.g.,][]{faisst2020}. Using the conversion factor from \citet{kennicutt1998}, this yields an obscured star formation rate (SFR) of $\mathrm{SFR}_\mathrm{IR} = (1.1 \times 10^{-10}) \times (L_\mathrm{IR}/L_\odot) = 268_{-55}^{+66}\,M_\odot\,\mathrm{yr}^{-1}$.\footnote{We convert the \citet{kennicutt1998} calibration to the Chabrier IMF assumed in this work by multiplying by $0.63$, following \citet{madau2014}.} The total SFR of HZ10, including also the UV-based SFR from \citet{capak2015}, is then $\mathrm{SFR} = 304_{-55}^{+67}\,M_\odot\,\mathrm{yr}^{-1}$. This suggests that  $f_\mathrm{obs} = 88_{-3}^{+2}\%$ of its star formation is obscured by dust, in line with the obscured fraction inferred for similarly massive galaxies at lower redshift \citep[$z\lesssim2.5$;][]{whitaker2017}, but in excess of what is typically seen in galaxies at $z\approx4-7$ \citep{fudamoto2020,algera2023}. \medskip

Given the robust sampling of the dust SED of HZ10 at six distinct wavelengths, it is possible to fit its far-IR emission with more complicated models than an optically thin modified blackbody. We discuss these additional fits in detail in Appendix \ref{app:dustFits}, while we summarize the results here. First of all, previous studies of HZ10 have generally assumed its dust emission to be optically thick at wavelengths shorter than $\lambda_\mathrm{thick} = 100\,\mu\mathrm{m}$ \citep{faisst2020,villanueva2024}. In the upper right panel of Fig.\ \ref{fig:dustFitsComparison}, we present a similar fit with $\lambda_\mathrm{thick}$ fixed to this value, which yields $T_\mathrm{dust}=51_{-7}^{+8}\,\mathrm{K}$, fully consistent with that measured in \citet{faisst2020} and \citet{villanueva2024}. The corresponding dust mass of $\log(M_\mathrm{dust}/M_\odot) = 7.8 \pm 0.2$ is furthermore consistent with our fiducial fit within the uncertainties.

Leveraging the new Band 10 observations, we can also attempt to constrain the wavelength where the optical depth reaches unity directly. We therefore perform another fit to the far-IR emission of HZ10 (lower left panel in Fig.\ \ref{fig:dustFitsComparison}) with $\lambda_\mathrm{thick}$ kept free. However, despite bracketing the peak of the dust SED with the Band 8, 9 and 10 data, the current observations cannot distinguish between mostly optically thin dust ($\lambda_\mathrm{thick} \ll 100\,\mu\mathrm{m}$) and dust with $\lambda_\mathrm{thick} \sim 100 - 150\,\mu\mathrm{m}$, while only large values of $\lambda_\mathrm{thick} \gtrsim 170\,\mu\mathrm{m}$ can currently be ruled out. While the dust temperature and mass cannot be well constrained in this general-opacity fit, we note that the inferred dust mass of $\log(M_\mathrm{dust}/M_\odot) = 7.9_{-0.5}^{+0.2}$ remains consistent with our fiducial optically thin value within the uncertainties.

Finally, we consider a multi-temperature dust model. HZ10 is nearly equally luminous in Bands 8, 9 and 10 (Table \ref{tab:properties}), which is characteristic of emission from dust at a range of temperatures \citep[e.g.,][]{magnelli2012b,casey2014}. Indeed, the spatially resolved analysis of HZ10 in \citet{villanueva2024}, which utilizes the high-resolution ALMA data in Bands 7 and 9, suggests $\Delta T_\mathrm{dust}\sim5-10\,\mathrm{K}$ differences between the dust temperatures of HZ10-C+E, HZ10-W, and the `bridge' connecting the two. To account for such dust at multiple temperatures, we perform a fit to the ALMA photometry of HZ10 adopting the multi-temperature prescription from \citet[][lower right panel in Fig.\ \ref{fig:dustFitsComparison}]{sommovigo2025}. This yields a dust mass of $\log(M_\mathrm{dust}/M_\odot) = 8.2_{-0.2}^{+0.3}$, which is fully consistent with the value inferred using our fiducial single-temperature model.

As is clear from both the brief above discussion and the more extended one in Appendix \ref{app:dustFits}, fitting more elaborate models to the ALMA photometry of HZ10 does not result in a significantly different inferred dust mass. Similarly, the infrared luminosity and dust emissivity index are not strongly dependent on the precise fitting function, while the dust temperature is more strongly affected by assumptions regarding the optical depth (see also \citealt{dacunha2021,ismail2023,algera2024b,lower2024}). For general consistency with much of the recent high-redshift literature \citep[e.g.,][]{bakx2021,sommovigo2022,algera2024a,algera2024b,valentino2024}, we proceed with our optically thin model as fiducial. Next-generation far-IR missions such as \textit{PRIMA} \citep{moullet2023} will be crucial to better constrain the importance of optically thick and multi-temperature dust at early cosmic epochs.

In what follows, we are mainly interested in the HZ10-C+E system, which \citet{telikova2024} found can be modeled as a disturbed rotating disk. Given that HZ10 is not resolved in the available Band 4 and 8 observations, we perform an additional fit to just the Band 7, 9 and 10 data. As in our fiducial model for HZ10, we assume a fully optically thin MBB, adopting a narrower Gaussian prior on $\beta_\mathrm{IR}$ with a mean (standard deviation) of $2.2$ ($0.4$) based on the global fit. We show the fit to HZ10-C+E in Figure \ref{fig:dataFit} and present the results in Table \ref{tab:mbb}. Although the Band 10 flux density of HZ10-C+E slightly exceeds our best-fit model, this does not significantly affect the fit owing to the low signal-to-noise ratio of the data. On the other hand, the high S/N of the Band 7 and 9 observations still enable accurate constraints on the dust mass of HZ10-C+E of $\log(M_\mathrm{dust}/M_\odot) = 7.9 \pm 0.2$. 

\begin{figure*}
    \centering
    \includegraphics[width=0.9\linewidth]{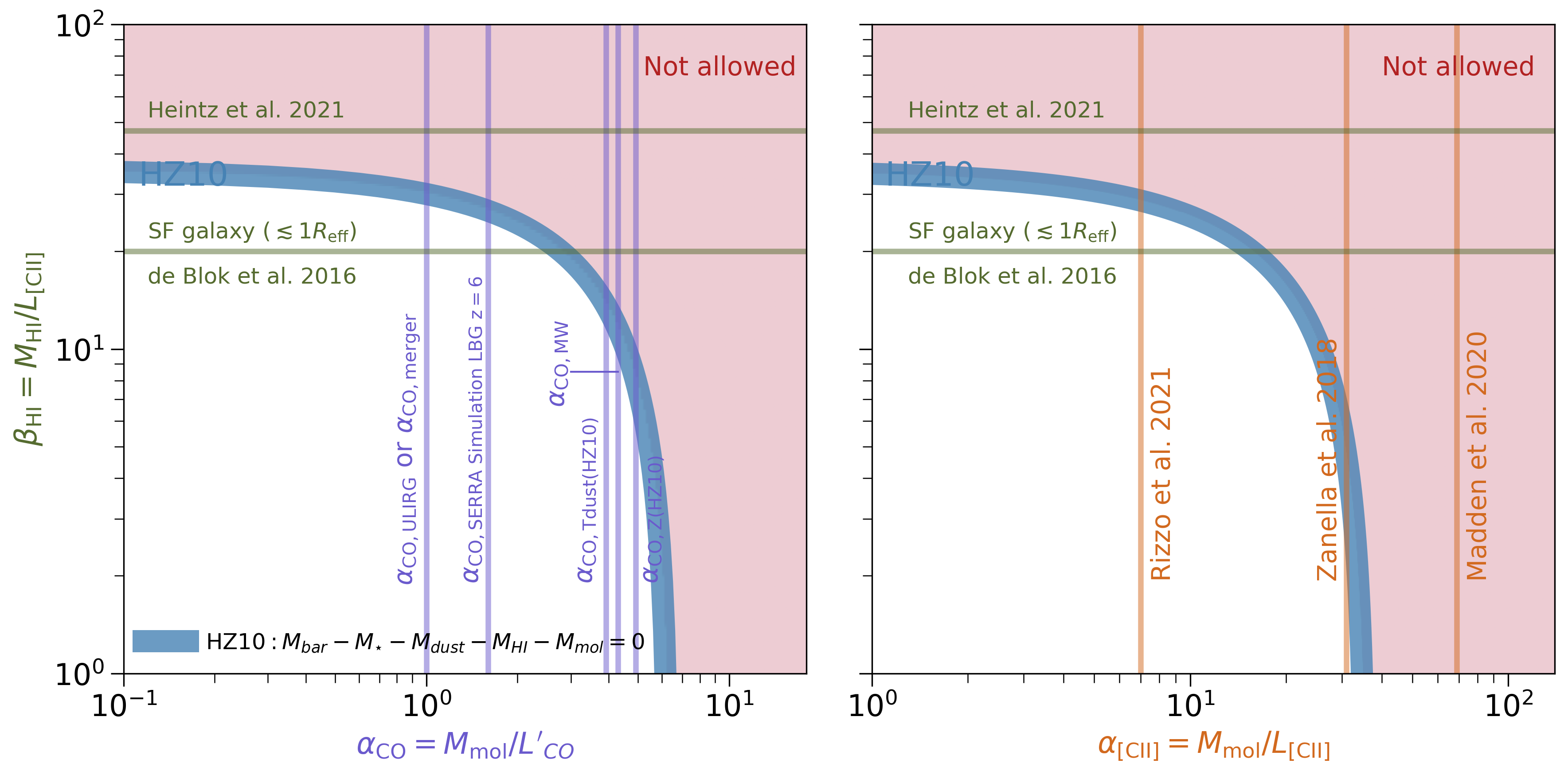}
    \caption{\textit{(Left)} Total mass budget of HZ10 as a function of the CO-to-H$_2$ ($\alpha_{\rm CO}$) and [CII]-to-HI ($\beta_{\rm HI}$) conversion factors. The blue line represents the case where $M_{\rm bar} - M_{\star} - M_{\rm dust} - M_{\rm HI} - M_{\rm mol} = 0$. The shaded regions indicate the forbidden region where this expression is less than zero. The green and red lines correspond to typical values of $\beta_{\rm HI}$ and $\alpha_{\rm CO}$, respectively, commonly assumed in the literature for different galaxy conditions. For the molecular gas, these include: MW-like $\alpha_{\rm CO}$ \citep[e.g.,][]{bolatto2013}, ULIRG-like or merger-like $\alpha_{\rm CO}$ \citep[e.g.,][]{downes1998,tacconi2008}, $\alpha_{\rm CO}$ from a \textsc{serra} simulated Lyman-break galaxy at $z=6$ \citep{vallini2018}, and $\alpha_{\rm CO}$ based on the dust temperature and metallicity of HZ10-C+E following \cite{magnelli2012} and \citet{bolatto2013}, respectively. For the atomic hydrogen gas, we include the calibration by \cite{heintz21} and the mass-to-light ratio observed within one effective radius in nearby star-forming galaxies \citep{deblok2016}. \textit{(Right)} Similar to the left panel, but here the x-axis represents the [CII]-to-H$_2$ ($\alpha_{\rm [CII]}$) conversion factor. The $\alpha_{\rm [CII]}$ calibrations by \cite{zanella18}, \cite{madden2020}, and \cite{rizzo21} are included.
    }
    \label{fig:MassBudget}
\end{figure*}

\section{Discussion}
\label{sec:discussion}

\subsection{The molecular and atomic gas mass budget in HZ10-C+E}
\label{sec:massBudget}

Given the wealth of ancillary data available for HZ10, we can study its mass budget in greater detail than for the large majority of main-sequence star-forming galaxies at $z\gtrsim4$. We begin by expressing the total mass of HZ10-C+E as:

\begin{equation}
M_{\rm bar} - M_{\star} - M_{\rm dust} - M_{\rm HI} - M_{\rm mol} = 0,
\end{equation}

\noindent where $M_{\rm bar}$ is the baryonic mass, $M_{\rm HI}$ is the atomic gas mass, and $M_{\rm mol}$ is the molecular gas mass.

We now discuss the values of each mass component listed in Eqs. (1)–(3) for HZ10-C+E. Based on a morpho-kinematic analysis of sensitive, high-angular-resolution ($\approx0.3\arcsec$) \cii\ data from the CRISTAL survey \citep{herrera-camus2025}, \citet{telikova2024} find that HZ10-C+E is a rotating disk. Its rotation was modeled using the Python-based forward modeling code {\sc DysmalPy} \citep[][]{davies2004,davies2004b,davies2011,cresci2009,wuyts2016,lang2017,price2021,lee2025}, yielding a low dark-matter fraction of $\approx30\%$ within the effective radius ($1\times R_{\rm eff}$) and a total baryonic mass of $\log(M_{\rm bar}/M_{\odot}) = 11.1^{+0.2}_{-0.3}$.

For the stellar mass, we have a global mass measurement based on SED modeling using the code CIGALE \citep{boquien2019} that corresponds to $\log(M_{\star}/M_{\odot}) = 10.35 \pm 0.37$. The details of the photometric filters and fitting parameters used in the modeling are provided in \citet[][Appendix B]{mitsuhashi2024}. Since \textit{JWST}/NIRCam data is not available for HZ10, we estimate that the stellar mass of HZ10-C+E is approximately $70\%$ of the global value, based on the scaling of the \textit{JWST}/NIRSpec 4~$\mu$m continuum flux between the HZ10-C+E and HZ10-W components \citep[][their Fig. 5]{jones2024}. 

For the atomic gas, the only available tracer is the \cii\ line, one of the primary coolants of the cold, neutral ISM \citep[e.g.,][]{wolfire2022}, and thus linked to the atomic gas in galaxies, particularly in their outer regions \citep[e.g.,][]{madden93,deblok2016}. %The atomic gas mass can be estimated as $M_{\rm HI} = \beta_{\rm HI} \times L_{\rm [CII]}$, where $\beta_{\rm HI}$ is the \cii-to-HI conversion factor. 
\citet{heintz21} provide a calibration of $\beta_{\rm HI}$ for star-forming galaxies as a function of metallicity based on direct measurements of relative abundances along lines of sight through the ISM of $\gamma$-ray burst host galaxies. This method provides an upper limit on the HI gas mass physically associated with the \cii-emitting region. For the metallicity of HZ10, this calibration yields $\beta_{\rm HI}\approx 47\, M_{\odot}/L_{\odot}$. If we aim to measure a $\beta_{\rm HI}$ value that is more physically representative of the star-forming disk --within $1-1.5\times R_{\rm eff}$, where the bulk of the CO and \cii\ line emission in HZ10 is detected-- we can adopt the mass-to-light ratio $M_{\rm HI}/L_{\rm [CII]}$ measured in nearby star-forming galaxies by \citet{deblok2016}. Within one effective radius ($R_{\rm eff}$), this ratio corresponds to $\beta_{\rm HI} \approx 20~M_{\odot}/L_{\odot}$, which is a factor of $\sim2$ lower than the value reported by \citet{heintz21}. Only at distances of $\approx3\times R_{\rm eff}$ does $\beta_{\rm HI}$ in nearby galaxies become comparable to that derived by \citet{heintz21}, reinforcing the interpretation that the latter calibration represents an upper limit to the total atomic gas content of the system \citep[see also][]{rowland2025_dla}. 

For the molecular gas, we use the detection of the CO(2-1) line \citep[$S_{\rm CO(2-1)}\Delta v=0.1\pm0.02$~Jy~km~s$^{-1}$; ][]{pavesi2019}. There are currently no observational constraints on the molecular gas excitation in normal star-forming galaxies at $z\approx5$. Therefore, we assume $L'_{\rm CO(2-1)}/L'_{\rm CO(1-0)} = 1$, consistent with the values observed in star-forming galaxies at $z\approx1$–2 \citep[e.g.,][]{aravena2014,daddi2015,genzel2015,tacconi20,boogaard20}. In both cases, we consider only the \cii\ or CO luminosity from HZ10-C+E. For the spatially resolved \cii\ observations, measuring the flux from the HZ10-C+E component is straightforward, as this region is spatially resolved and accounts for $\sim60\%$ of the total flux \citep[$S^{total}_{\rm [CII]}\Delta v=4.9\pm0.02$~Jy~km~s$^{-1}$; ][]{telikova2024}. However, for the CO(2–1) line, only a global measurement is available \citep{pavesi2019}. By analogy with the \cii\ case, we therefore assume that $\sim60\%$ of the total CO(2–1) emission originates from the HZ10-C+E component.

For the CO-to-H$2$ conversion factor, we consider four values: (1) a Milky Way (MW)–based value \citep[$\alpha_{\rm CO} = 4.3~{\rm M}_{\odot}~({\rm K~km~s}^{-1})^{-1}$; e.g.,][]{bolatto2013}; (2) the typical value assumed for nearby ULIRGs and merger-driven starbursts \citep[$\alpha_{\rm CO} = 1~{\rm M}_{\odot}~({\rm K~km~s}^{-1})^{-1}$; e.g.,][]{downes1998,tacconi2008}; (3) the $\alpha_{\rm CO} = 1.5~{\rm M}_{\odot}~({\rm K~km~s}^{-1})^{-1}$ obtained for a $z=6$ Lyman-break galaxy from the \textsc{serra} simulations with a metallicity similar to that of HZ10 \citep{vallini2018}; and (4-5) two estimates specific to HZ10-C+E: $\alpha_{\rm CO} = 3.9~{\rm M}_{\odot}~({\rm K~km~s}^{-1})^{-1}$, derived from its dust temperature following the calibration by \citet{magnelli2012}, and $\alpha_{\rm CO} = 4.9~{\rm M}_{\odot}~({\rm K~km~s}^{-1})^{-1}$, estimated from its metallicity following \citet{bolatto2013}.

Alternatively to CO, we can use the \cii\ luminosity as a tracer of the molecular gas. For the \cii-to-H$_2$ conversion factor, we adopt three calibrations: (1) the calibration by \citet{zanella18}, based on (U)LIRGs and main-sequence star-forming galaxies up to $z\sim2$; (2) the calibration by \citet{madden2020}, derived from local dwarf galaxies; and (3) the calibration by \citet{rizzo21}, based on the dynamical properties of 5 dusty star-forming galaxies at $z\approx4.5$, of which 4 lie above the main-sequence of star-forming galaxies, as opposed to HZ10.

Figure~\ref{fig:MassBudget} shows the total mass budget of HZ10 as a function of $\alpha_{\rm CO}$ (left panel), $\alpha_{\rm [CII]}$ (right panel), and $\beta_{\rm HI}$. The blue line represents the solution to Equation~1, while the red shading indicates the forbidden region where $M_{\rm bar} < M_{\star} + M_{\rm dust} + M_{\rm mol} + M_{\rm HI}$. In both panels, the values of $\beta_{\rm HI}$ from \citet{heintz21}, derived using the metallicity of HZ10, and from the inner $\sim1\times R_{\rm eff}$ of nearby galaxies \citep{deblok2016} are shown as green horizontal lines. 

We begin by discussing the left panel, where the molecular gas mass is derived from the CO line emission. The $\alpha_{\rm CO}$ conversion factors inferred from the dust temperature and metallicity of HZ10–C+E are both comparable to the Milky Way–like value of $\alpha_{\rm CO}$. Adopting this conversion factor yields a molecular gas mass of $M_{\rm mol}\approx 6.0\times10^{10}~{\rm M}_{\odot}$ and a molecular gas fraction, defined as $\mu_{\rm mol}=M_{\rm mol}/M_{\star}$, of $\mu_{\rm mol}\approx 3.8$, which is probably too gas rich compared to the expectations for star-forming galaxies at this redshift. If instead we assume the $\alpha_{\rm CO}$ value typical of (U)LIRGs, the inferred molecular gas fraction decreases by a factor $\alpha_{\rm CO}/\alpha_{\rm ULIRG}=4.3/1$, that is, to $\mu_{\rm mol}\approx 0.9$. This value is more in line with the expectations from scaling relations of galaxies between $0\lesssim z \lesssim 4$ \citep[e.g.,][]{tacconi20}. Interestingly, a similarly low $\alpha_{\rm CO}$ value has been reported by \citet{vallini2018} for a high-redshift star-forming galaxy drawn from the {\sc{serra}} simulations, with a metallicity comparable to that of HZ10. In particular, \citet{vallini2018} found $\alpha_{\rm CO}\approx1.5$, attributing this lower value to the combination of strong turbulence, high gas densities, and the elevated CMB temperature at early cosmic times.

It is moreover important to note that when a Milky Way-based (or similar) $\alpha_{\rm CO}$ is adopted, the resulting mass budget leaves no room for the atomic gas mass estimated using the calibration by \citet{heintz21}, as the intersection of the corresponding lines lies within the forbidden region. Conversely, the only scenario in which the CO-based molecular gas and the H\,I mass are consistent with the total mass budget of HZ10 is when the \cii-to-$M_{\rm HI}$ calibration factor based on the inner $R_{\rm eff}$ radius of nearby galaxies, $\beta_{\rm HI} \approx 20 M_{\odot}/L_{\odot}$, is adopted. \\

We continue our analysis, this time considering the \cii-based molecular gas mass estimates (right panel of Fig.~\ref{fig:MassBudget}). Using the calibration by \citet{rizzo21}, we derive a molecular gas mass of $M_{\rm mol} = 1.7\times10^{10}\,M_{\odot}$ and a corresponding molecular gas fraction of $\mu_{\rm mol} = 0.5$. The other three \cii–to–H$_2$ calibrations yield systematically higher molecular gas masses: the \citet{zanella18} and \citet{madden2020} relations give values larger by factors of $\approx4$ and $\approx10$, respectively.

Similar to the case of the CO-based molecular gas estimates, and considering the associated uncertainties, in the case of HZ10-C+E only the calibration by \citet{rizzo21} can be used together with the HI mass calibration by \citet{heintz21} without exceeding the total dynamical mass of the system. Conversely, the calibration by \citet{zanella18} is consistent only when adopting the \cii-to-$M_{\rm HI}$ conversion factor derived for the inner $R_{\rm eff}$ of nearby galaxies. Finally, the calibration by \citet{madden2020}, based on a sample of low-metallicity dwarf galaxies, significantly overpredicts the molecular gas mass relative to the other \cii- and CO-based tracers. This is expected, as the ISM conditions in low-metallicity dwarfs most likely differ from those in massive, dusty, star-forming galaxies at high redshift such as HZ10.

All in all, this simple analysis shows the difficulties of simultaneously measuring the H\,I and H$_2$ masses from \cii\ line emission, given the varied nature of the calibrations and the fact that \cii\ emission is multi-phase, arising from both molecular and atomic gas. A more practical and physically motivated approach moving forward may be to use the \cii\ line as a tracer of the total cold gas -- that is, the combined molecular and atomic components of the ISM. This is analogous to the approach commonly adopted when using the dust continuum as a gas tracer, since dust is mixed with both the molecular and atomic phases of the ISM \citep[e.g.,][]{scoville17}. This perspective has also been followed in recent studies, such as the \textsc{serra} simulations by \citet{vallini25}, which derived scaling relations between $M_{\rm gas}$ and the \cii\ line luminosity that depend on both the metallicity and the size of the system. Following this approach, in the case of HZ10, the \cii\ conversion factor between the total cold gas mass ($M_{\rm ISM}=M_{\rm HI}+M_{\rm mol} = M_{\rm bar}-M_{\star} - M_\mathrm{dust}$) and the \cii\ luminosity is $\alpha_{\rm [CII]}^{\rm ISM} = 39^{+50}_{-25}~M_{\odot}/L_{\odot}$. 

One clear outcome of this analysis --independent of the uncertainties in the baryonic and stellar mass estimates and in the adopted calibration factors-- is that HZ10 is a galaxy rich in cold ISM gas, with a cold gas-to-stellar mass fraction of $\approx 2$. This high value is consistent with the \cii-based measurements for star-forming galaxies at $4\lesssim z \lesssim7$ reported by \citet{dessaugeszavadsky2020} and \citet{aravena2024}, as well as with the extrapolation of the scaling relation derived by \citet{tacconi2018,tacconi20} for star-forming galaxies at $z\lesssim4$, and with the \textsc{serra} simulated galaxies presented by \citet{vallini25}.

The original \textsc{serra} relation for $\log \alpha_{\rm [CII]}$ versus metallicity (see Eq.~1 in \citealt{vallini25}) was calibrated using metallicities averaged over a $5\times5~\mathrm{kpc}^2$ region around each simulated galaxy. This makes it not directly comparable to our derivation for HZ10. To address this, we recomputed the relation using the metallicity averaged within the [C\,\textsc{ii}] half-light radius of the simulated galaxies, ensuring a more consistent spatial comparison (Vallini et al. in preparation). This yields a relation of
\[
\log \alpha_{\rm [CII]} = -0.38\,\log(Z/Z_{\odot}) + 1.12,
\]
which corresponds to $\alpha_{\rm [CII]} \approx 16$ for HZ10-C+E. This value is consistent, within the uncertainties, with the dynamically inferred value.

\subsection{The dust-to-gas and dust-to-metal ratios of HZ10 and HZ10-C+E}
\label{sec:dustBuildUp}

\begin{figure*}
    \centering
    \includegraphics[width=0.8\linewidth]{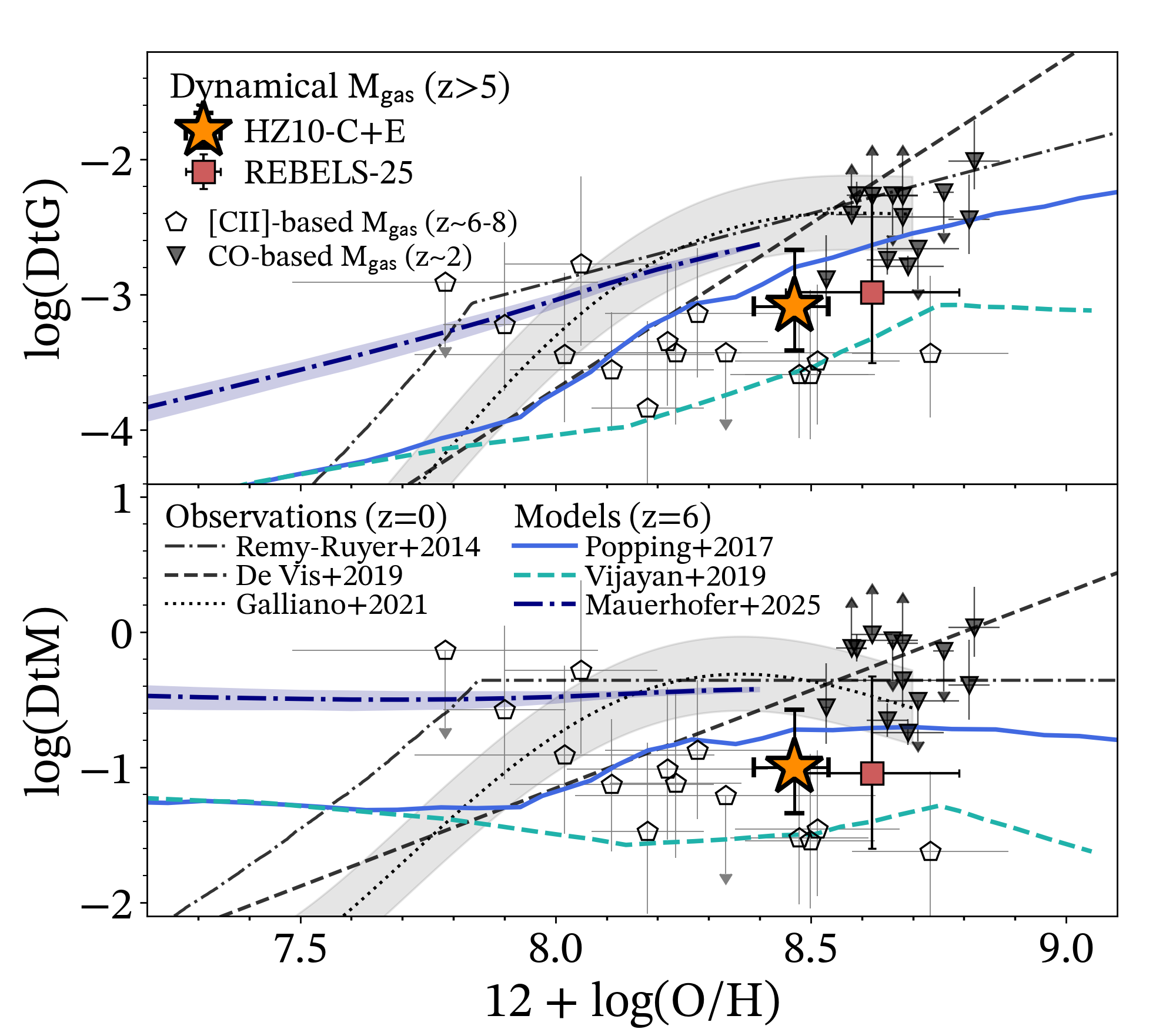}
    \caption{The dust-to-gas (DtG; \textit{upper panel}) and dust-to-metal (DtM; \textit{bottom panel}) ratios of HZ10-C+E (orange star) as a function of metallicity. For the dust mass, we use our fiducial optically thin model, while the gas mass of HZ10-C+E is determined as the difference of its baryonic mass -- inferred through kinematic modeling by \citet{telikova2024} -- and stellar mass \citep{mitsuhashi2024}. The metallicity of HZ10-C+E was recently inferred through \textit{JWST}/NIRSpec IFU observations by \citet{jones2024}. We compare to a similar dynamically-inferred DtG and DtM for the $z=7.31$ galaxy REBELS-25 \citep[red square;][]{rowland2024,rowland2025,algera2025}, while at Cosmic Noon we leverage ALMA-based dust and gas mass measurements from \citet{shapley2020} and \citet[][black triangles]{popping2023}. The open pentagons represent \cii{}-based measurements from \citet{algera2025}. Finally, we overlay several local relations \citep[][grey lines]{remyruyer2014,devis2019,galliano2021} as well as model tracks at $z\approx6$ \citep[][colored lines]{popping2017,vijayan2019,mauerhofer2025}. Despite its metal-enriched nature, we find that HZ10-C+E falls below the local scaling relations, suggesting either a limited efficiency of ISM dust growth in HZ10, or efficient dust destruction.}
    \label{fig:dustBuildUp}
\end{figure*}

Given that our ALMA observations both bracket the peak of the dust SED and cover the Rayleigh-Jeans tail, we obtain robust constraints on the total dust mass of HZ10 of $\log(M_\mathrm{dust}/M_\odot) = 8.0 \pm 0.1$ given our fiducial optically thin model. We also constrain its dust emissivity index to be $\beta_\mathrm{IR}=2.2\pm0.4$, which is consistent with what is typically seen in galaxies at Cosmic Noon \citep{dacunha2021,liao2024,bendo2025} as well as at higher redshifts \citep{witstok2023,algera2024b}.

Using the dust mass obtained from our SED-fitting and the stellar mass from \citet{mitsuhashi2024}, we infer a dust-to-stellar mass ratio of $\log(M_\mathrm{dust}/M_\star) = -2.31 \pm 0.39$ for HZ10, equivalent to $M_\mathrm{dust}/M_\star = 5_{-3}^{+7} \times 10^{-3}$. This value is consistent with typical dust-to-stellar mass ratios of $z\approx4-8$ star-forming galaxies \citep{bakx2021,sommovigo2022,witstok2023,algera2024b,algera2025,sawant2025}. Assuming a \citet{chabrier2003} IMF and the \citet{michalowski2015} framework, the measured ratio for HZ10 corresponds to an IMF-averaged supernova dust yield of $y_\mathrm{SN} = 0.41_{-0.24}^{+0.60}\,M_\odot/\mathrm{SN}$. This is in line with the large range of values suggested for SNe ranging from $y_\mathrm{SN} < 10^{-3}\,M_\odot$ to $y_\mathrm{SN} \sim 1\,M_\odot$, as recently compiled by \citet{schneider2024}.

However, converting the dust-to-stellar mass ratio to a SN dust yield implicitly assumes negligible contribution from ISM dust growth and AGB stars, which even at high redshift is not necessarily the case \citep[e.g.,][]{lesniewska2019,algera2024b,algera2025}. To better understand the pathways of dust build-up in HZ10, we shift our focus to HZ10-C+E, where we can leverage the kinematic constraints on its baryonic mass. As in the previous section, we determine the total (atomic + molecular) gas mass of HZ10+C-E as $M_\mathrm{gas} = M_\mathrm{mol} + M_\mathrm{HI} = M_\mathrm{bar} - M_\star - M_\mathrm{dust}$. Using this gas mass, we compute the dust-to-gas ratio $\mathrm{DtG} = M_\mathrm{dust} / M_\mathrm{gas}$ and -- combined with the metallicity of $Z\sim0.6\,Z_\odot$ from \citet{jones2024} -- the dust-to-metal ratio $\mathrm{DtM} = \mathrm{DtG} / Z$.

We infer $\log(\mathrm{DtG}) = -3.08_{-0.33}^{+0.40}$ and $\log(\mathrm{DtM}) = -1.00_{-0.34}^{+0.41}$ for HZ10-C+E, and show its DtG and DtM as a function of metallicity in Fig.\ \ref{fig:dustBuildUp}. To place HZ10+C-E in context, we compare to a variety of studies spanning $z\approx0-8$. First of all, we compare to several scaling relations between DtG, DtM and metallicity derived for local galaxies, adopting the best-fit trends from \citet{remyruyer2014, devis2019} and \citet{galliano2021}. These works typically infer dust masses from detections in various \textit{Herschel} bands, while gas masses include both molecular and atomic hydrogen.

At slightly higher redshift ($z\approx1.5 - 2.5$), we compare to the works by \citet{shapley2020} and \citet{popping2023}. Both studies combine Keck/MOSFIRE rest-optical spectra with ALMA-based dust and gas masses, the latter obtained from low- and mid-$J$ CO lines. As such, these works do not explicitly account for atomic hydrogen, and their DtG and DtM ratios can therefore be considered upper limits. \citet{shapley2020} adopt a fixed $\alpha_\mathrm{CO} = 3.6~{\rm M}_{\odot}~({\rm K~km~s}^{-1})^{-1}$, while \citet{popping2023} use the metallicity-dependent prescription from \citet{accurso17}, which ranges from $\alpha_\mathrm{CO} \sim 2.7 - 8.1~{\rm M}_{\odot}~({\rm K~km~s}^{-1})^{-1}$ for their sample. The dust mass estimates in both works are based on the assumption of a fixed dust temperature of $T_\mathrm{dust}\sim25 - 35\,\mathrm{K}$. We scale their dust masses to the same dust model adopted here (i.e., the $\kappa_{850}$ from \citealt{dunne2000}; Section \ref{sec:results}), which reduces the \citet{shapley2020} and \citet{popping2023} dust masses by $0.19$ and $0.32\,\mathrm{dex}$, respectively.

At higher redshifts ($z\approx6-8$), we compare to \citet{algera2025}, who recently compiled the dust-to-gas and dust-to-metal ratios of $15$ galaxies with ALMA \ciil{} observations, predominantly taken from the REBELS-IFU survey \citep[][]{bouwens2022,rowland2025}. \citet{algera2025} adopt the \citet{zanella18} conversion to infer the gas masses of their sample, which they argue accounts for both molecular and atomic gas \citep[see also][]{dessaugeszavadsky2020,aravena2024}. This moreover agrees with our analysis in Section \ref{sec:massBudget}, where applying the \citeauthor{zanella18} \cii{}-to-$M_\mathrm{mol}$ conversion factor indeed leaves little room for atomic gas in HZ10-C+E. To match our dust model, we scale the \citet{algera2025} dust masses downwards by $0.33\,\mathrm{dex}$.

We also compare to the $z=7.31$ galaxy REBELS-25, which \citet{rowland2024} recently demonstrated to be a dynamically cold disk based on high-resolution \cii{} observations. Leveraging their dynamical mass measurement, as well as the stellar mass reported in \citet{rowland2025} and the dust mass inferred by \citet{algera2024b}, we infer the total gas mass of REBELS-25 in the same way as for HZ10-C+E in Section \ref{sec:massBudget}. Given that \citet{rowland2025} quote only the dynamical mass of REBELS-25, we adopt a dark matter fraction of $30\,\%$ -- the same as inferred for HZ10-C+E by \citet{telikova2024} -- to convert this to a baryonic mass. We moreover adopt its metallicity to be $12+\log(\mathrm{O/H}) = 8.62 \pm 0.17$ as inferred by \citet{rowland2025}.

Finally, we compare to three model tracks from the semi-analytical models by \citet{popping2017}, \citet{vijayan2019} and \citet{mauerhofer2025}, evaluated at $z\approx6$. These models include both dust production and destruction processes, albeit with different efficiencies. Among the three, the \citet{popping2017} models predict the largest dust enrichment by $z\approx6$, with the DtG ratios approaching those observed in local galaxies. In their fiducial model, dust production takes place primarily through growth in the ISM, which is predicted to be particularly efficient for massive and metal-rich galaxies such as HZ10.

The \citet{vijayan2019} models similarly predict that ISM growth is the dominant dust production mechanism in galaxies at $z\approx6$, although it proceeds at a lower rate than in the models by \citet{popping2017}. As a result, the DtG and DtM ratios predicted by the \citet{vijayan2019} models are $\sim1\,\mathrm{dex}$ lower than those by \citet{popping2017} at high metallicities ($12 + \log(\mathrm{O/H}) \gtrsim8.2$). Furthermore, due to this delayed grain growth, the \citet{vijayan2019} models predict a more or less flat trend between dust-to-metal ratio and metallicity at $z\approx6$, while only by $z\lesssim3$, the DtM ratio starts to consistently increase for more metal-rich galaxies.

Finally, we compare to the recent semi-analytical models by \citet{mauerhofer2025}, which build upon the dust production and destruction framework from \citet{dayal2022}. In their models, dust production is predominantly due to supernovae, which are assigned a fixed dust yield of $y_\mathrm{SN}=0.5\,M_\odot$. This results in a linear scaling between DtG ratio and metallicity, while the dust-to-metal ratio is independent of metallicity. While the \citet{mauerhofer2025} models are able to produce massive galaxies by $z\approx6$ (up to $M_\star \sim 10^{11.5}\,M_\odot$), they are predicted to, on average, have relatively low metallicities ($Z\lesssim0.2\,Z_\odot$). Consequently, the \citet{mauerhofer2025} models do not extend beyond $Z \gtrsim 0.5\,Z_\odot$ by $z\approx6$, slightly below the metallicity of HZ10. Nevertheless, owing to the high SN dust yields implemented in their models, \citet{mauerhofer2025} predict the highest DtG and DtM ratios at a fixed metallicity among the three models we compare to. \\

Overall, we find that the DtG and DtM ratios of HZ10-C+E fall $\sim0.5 - 1\,\mathrm{dex}$ below the local scaling relations. Given its high metallicity ($Z\sim0.6\,Z_\odot$; \citealt{jones2024}), HZ10 is expected to lie in the regime where ISM dust growth is efficient \citep[$Z\gtrsim0.2\,Z_\odot$; e.g.,][]{asano2013,choban2024}. The lower DtG and DtM ratios may therefore suggest that HZ10 has not yet efficiently formed dust through ISM dust growth, indicating this process operates on a longer characteristic timescale. Alternatively, given that HZ10 is undergoing a powerful starburst, it is possible that the resulting supernovae are able to efficiently destroy dust within the galaxy through shattering and sputtering, or eject it from the galaxy through outflows.

The other high-redshift galaxy with dynamical constraints on its total gas mass and multi-band constraints on $M_\mathrm{dust}$, REBELS-25, shows a similar DtG and DtM ratio to HZ10-C+E. Unlike HZ10, REBELS-25 does not appear to be undergoing a merger \citep{rowland2024}, and it has a $\sim3\times$ lower star formation rate than HZ10 \citep{algera2024b}. Both galaxies show a typical DtM ratio of $\log(\mathrm{DtM}) \approx -1$, which is consistent with the values obtained for $z\sim6-8$ galaxies by \citet{algera2025} leveraging \cii{}-based gas masses, but lower than what is seen in galaxies in the nearby Universe.

The measured DtG and DtM ratios of HZ10-C+E place it in between the model tracks from \citet{popping2017} and \citet{vijayan2019}, which we recall ascribe most of dust production at $z\approx6$ to growth in the ISM. Consequently, this would suggest that grain growth is already moderately efficient at $z\approx6$. We caution, however, that \citet{popping2017} also explore a variety of alternative dust production models in their work, and demonstrate that efficient SN dust production can also lead to similar DtG and DtM ratios as in their fiducial model shown in Figure \ref{fig:dustBuildUp}. Distinguishing between these scenarios requires a significantly larger sample of high-redshift galaxies with robust dust and gas mass measurements -- especially at the metal-poor end where models relying on ISM dust growth predict lower DtG ratios. Given the intrinsic faintness of such metal-poor and therefore typically low-mass galaxies, exploiting gravitational lensing provides a promising pathway of studying dust build-up in such systems \citep[c.f.,][]{heintz2023,heintz2025,valentino2024}.

We find that the dust-to-gas and dust-to-metal ratios of both HZ10-C+E and REBELS-25 are inconsistent with those predicted by the \citet{mauerhofer2025} models, which exceed the measurements by $\sim0.5\,\mathrm{dex}$. This is particularly interesting given that \citet{mauerhofer2025} demonstrate that their modeled dust-to-stellar mass ratios match those observed in high-redshift galaxies from ALPINE \citep{fudamoto2020} and REBELS \citep{bouwens2022,inami2022}. In turn, it is imperative to explore a wider range of scaling relations -- beyond the typical comparison of dust and stellar masses \citep[e.g.,][]{lesniewska2019,witstok2023,algera2024a} -- to more clearly distinguish between various models of dust build-up.

\section{Conclusions}
\label{sec:conclusions}

In this paper, we have studied the ISM properties of the $z=5.65$ main-sequence, star-forming galaxy HZ10 in detail. In particular, we focus on the key question of how much gas (atomic and molecular) and dust is present in its ISM. To address this, we combine ALMA dust continuum observations \citep{faisst2020,villanueva2024} including new Band 10 and 4 observations, %covering the wavelength range $\lambda_{\rm rest} \approx 50$--$300$~$\mu$m, 
detailed kinematics from \cii\ line observations \citep{telikova2024}, \textit{JWST}/NIRSpec measurements of the metallicity \citep{jones2024}, and JVLA observations of the molecular gas via the CO(2-1) transition \citep{pavesi2019}. Our main results can be summarized as follows:

\begin{enumerate}
    \item We detect the dust continuum emission in ALMA Bands 10 and 4, probing the peak and Rayleigh–Jeans tail of the dust SED, respectively. Combined with ancillary ALMA data \citep{faisst2020,villanueva2024}, this allows us to cover the dust SED from $\lambda_{\rm rest} \approx 50$--$300$~$\mu$m, providing the most complete rest-frame IR SED for a normal star-forming galaxy at this redshift. %The dust SED peak is broad, covering the range $\lambda_{\rm rest} \approx 50$ to $80~\mu$m, which is characteristic of emission from dust at a range of temperatures \citep[e.g.,][]{casey2014}.
    \medskip
    \item Fitting the global dust SED of HZ10 with a single-temperature modified blackbody and assuming optically thin emission, we infer a dust mass of $\log(M_\mathrm{dust} / M_\odot) = 8.0 \pm 0.1$, a dust temperature of $T_\mathrm{dust} = 37_{-5}^{+6}\,\mathrm{K}$, and a dust emissivity index of $\beta_\mathrm{IR} = 2.2 \pm 0.4$. This corresponds to a total infrared luminosity of $\log(L_\mathrm{IR}/L_\odot) = 12.4 \pm 0.1$, and a total (UV+IR) star formation rate of $\mathrm{SFR} = 304_{-55}^{+67}\,M_\odot\,\mathrm{yr}^{-1}$. We infer consistent values when fitting more sophisticated dust models to the ALMA data.

\end{enumerate}

\noindent We proceed by focusing on HZ10-C+E, the rotationally-supported component of the HZ10 system \citep{telikova2024} for which we can infer the total ISM mass based on its \cii{} kinematics. Our results can be summarized as follows:

\begin{enumerate}[resume]

    % SED fitting
    \item Regarding the independent calibrations of the molecular and atomic gas masses based on the \cii\ line emission reported in the literature, we find that, within the uncertainties, the total gas mass is compatible with the total mass budget from the kinematic modeling only if: (1) the $\beta_{\rm HI}$ calibration by \citet{heintz21} is combined with a typical $\alpha_{\rm CO}$ for (U)LIRGs or with the $\alpha_{\rm [CII]}$ values from \citet{rizzo21}; or (2) the $\beta_{\rm HI}$ value found within $1\,R_{\rm eff}$ in nearby galaxies by \citet{deblok2016} is combined with either the Milky Way $\alpha_{\rm CO}$ or the $\alpha_{\rm [CII]}$ calibration from \citet{zanella18}. In this sense, because we do not know a priori what fraction of the \cii\ emission in high-redshift galaxies arises from the molecular or the atomic phase, it is preferable to adopt a calibration factor that directly targets the total (atomic + molecular) gas mass, as also advocated by the \textsc{serra} simulation work of \citet{vallini25}. For HZ10-C+E, we find that the \cii-to-$M_{\rm ISM}$ conversion factor corresponds to $\alpha_{\rm [CII]}^{\rm ISM} = 39^{+50}_{-25}~{\rm M}_{\odot}\,L_{\odot}^{-1}$.
    
    \medskip
    % atomic and molecular mass
    \item Combining the total gas mass obtained from the \cii{} kinematics of HZ10-C+E with the dust mass from SED fitting ($\log(M_\mathrm{dust}/M_\odot) = 7.9 \pm 0.2$), we study its dust production pathways through the dust-to-gas ($\log\mathrm{DtG} = -3.08_{-0.33}^{+0.40}$) and dust-to-metal ($\log\mathrm{DtM}=-1.00_{-0.34}^{+0.41}$) ratios. We find that HZ10-C+E falls below the local scaling relations of DtG and DtM with metallicity, suggesting less efficient dust production or enhanced dust destruction compared to local galaxies. Aided by a comparison to theoretical models at $z\approx6$, we speculate the lower dust contents are due to inefficient grain growth in its ISM, despite the relatively metal-enriched nature of HZ10-C+E ($Z\approx0.6\,Z_\odot$; \citealt{jones2024}).
\end{enumerate}

Using HZ10 as a case study, we have demonstrated how synergetic observations from some of the world's most advanced telescopes (\textit{JWST}, ALMA, JVLA) can be used to unravel the detailed baryonic mass budget of high-redshift galaxies. In the future, similar observations of larger galaxy samples will be able to robustly constrain the various conversion factors between emission line luminosities and molecular/atomic gas masses (i.e., $\alpha_\mathrm{CO}, \alpha_\text{\cii{}}, \beta_\mathrm{HI}$), as well as constrain their dust contents through multi-band far-IR continuum sampling. Combined, such observations will provide the most detailed possible insights into the ISM contents and evolutionary pathways of the distant galaxy population.

\begin{acknowledgements}
      H.S.B.A gratefully acknowledges support from Academia Sinica through grant AS-PD-1141-M01-2. R.H.-C. thanks the Max Planck Society for support under the Partner Group project "The Baryon Cycle in Galaxies" between the Max Planck for Extraterrestrial Physics and the Universidad de Concepción. R.H-C. also gratefully acknowledges financial support from ANID - MILENIO - NCN2024\_112 and ANID BASAL FB210003. M.A. is supported by FONDECYT grant number 1252054, and gratefully acknowledges support from ANID Basal Project FB210003 and ANID MILENIO NCN2024\_112. E.d.C. acknowledges support from the Australian Research Council (projects DP240100589). P.D. warmly acknowledges support from an NSERC discovery grant (RGPIN-2025-06182). G.C.J. acknowledges support by the Science and Technology Facilities Council (STFC), by the ERC through Advanced Grant 695671 ``QUENCH'', and by the UKRI Frontier Research grant RISEandFALL.
\end{acknowledgements}

\bibliographystyle{aa}
\bibliography{main}

\begin{appendix}
\section{Multi-band ALMA cutouts of HZ10}
\label{app:cutouts}

\begin{figure}
    \centering
    \includegraphics[width=1.0\linewidth]{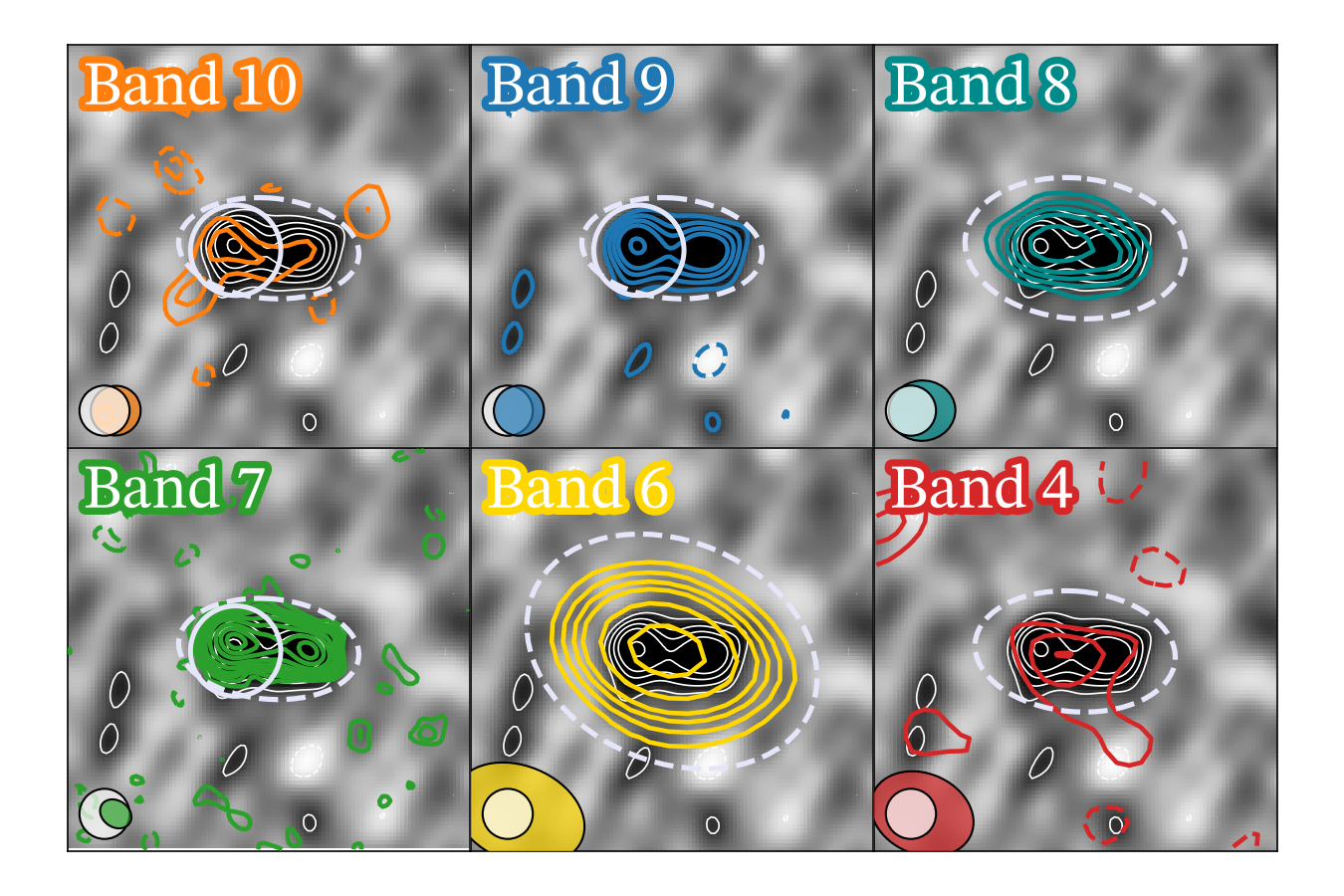}
    \caption{Six-band ALMA continuum maps of HZ10 ($4''\times4''$). Contours are drawn at $2,3,4,5,7$ and $10\sigma$, after which they continue in steps of $5\sigma$. Negative contours are dashed. The synthesized beam of each observation is shown as a filled ellipse, and the Band 9 contours and synthesized beams are shown in white in each panel for reference. The apertures used for extracting the flux density of the full HZ10 system and -- where possible -- that of HZ10-C+E are shown as dashed and solid lavender regions, respectively.}
    \label{fig:allCutouts}
\end{figure}

We show the ALMA continuum images of HZ10 in Bands 10 through 4 in Figure \ref{fig:allCutouts}. The Band 9 data were previously presented in \citet{villanueva2024}. The Band 8 and 6 observations were presented by \citet{faisst2020}, while the Band 7 observations are also shown in \citet{telikova2024} and \citet[][here HZ10 is known as CRISTAL-22]{herrera-camus2025}. The Band 10 and 4 observations are presented for the first time in this work.

\section{Physical properties of HZ10 and HZ10-C+E}
\label{app:properties}

We present the multi-band ALMA flux measurements and other relevant physical parameters of HZ10 and HZ10-C+E in Table \ref{tab:properties}.

\begin{table*}[t]
\caption{ALMA continuum flux densities and relevant physical parameters of HZ10 and HZ10-C+E. }
%\centering
%\resizebox{\linewidth}{!}{%
\begin{tabular}{l|c|c}
\hline\hline
& \textbf{HZ10} & \textbf{HZ10-C+E} \\
\hline

$\lambda_\mathrm{obs}$ [$\mu$m] & Flux [$\mu$Jy] & Flux [$\mu$Jy] \\
\hline

$2061$ (Band 4) & $101 \pm 42$ & - \\
$1318$ (Band 6) & $735 \pm 98$ & - \\
$1026$ (Band 7) & $1529 \pm 56$ & $883 \pm 37$ \\
$738$ (Band 8) & $2766 \pm 414$ & - \\
$447$ (Band 9) & $3484 \pm 487$ & $1831 \pm 326$ \\
$348$ (Band 10) & $3865 \pm 1317$ & $2797 \pm 883$ \\
\hline \hline

$z_\text{\cii{}}^{a}$ & $5.653$ & $5.653$ \\ 
$L_\text{\cii{}}/(10^9\,L_\odot)^{a}$ & $5.88 \pm 0.25$ & $3.51 \pm 0.20$ \\
$L_\text{CO(2-1)}/(10^{10}\,\mathrm{K\,km\,s}^{-1}\mathrm{pc}^{2})^{b}$ & $2.9 \pm 0.6$ & $(\mathit{1.7 \pm 0.4})$ \\
$\log(M_\star/M_\odot)^{c}$ & $10.35 \pm 0.37$ & $(\mathit{10.20 \pm 0.22})$ \\
$Z/Z_\odot^{d}$ & $0.53 - 0.71$ $(\mathit{0.60 \pm 0.10})$ & $0.53 - 0.63$ $(\mathit{0.60 \pm 0.10})$ \\

\hline\hline 
\end{tabular}
%}
\\\small\textbf{Notes:} Uncertainties on the flux densities are reported without the added calibration error (Section \ref{sec:data}). When a range of measurements or no resolved measurements are available, we denote our adopted fiducial values in parentheses and in italics. \\
$^{(a)}$ \cii{}-based redshift and luminosity from \citet{herrera-camus2025}. HZ10 and HZ10-C+E are denoted as CRISTAL-22 and CRISTAL-22a in their work, respectively. \\
$^{(b)}$ CO(2-1) luminosity measured by \citet{pavesi2019}. \\
$^{(c)}$ Stellar mass from \citet{mitsuhashi2024}. \\
$^{(d)}$ The metallicity of the three components of HZ10 were measured individually by \citet{jones2024}, and range from $Z/Z_\odot = 0.53 \pm 0.09$ for HZ10-E to $Z/Z_\odot = 0.71 \pm 0.12$ for HZ10-C. We adopt a fiducial $Z/Z_\odot = 0.60 \pm 0.10$ based on these values.
\label{tab:properties}
\end{table*}

\section{Additional dust SED fits of HZ10}
\label{app:dustFits}

\begin{figure*}
    \centering
    \includegraphics[width=0.9\linewidth]{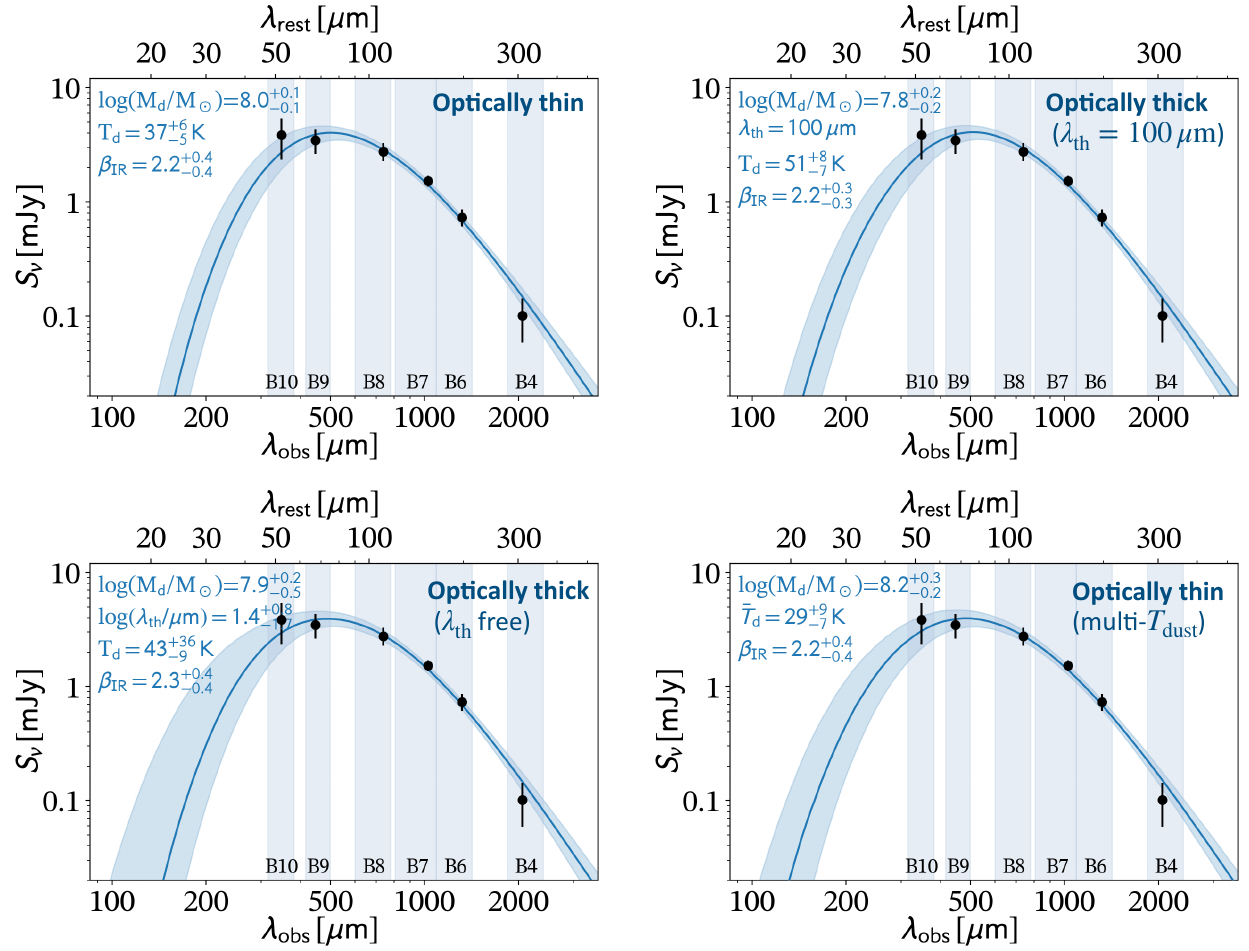}
    \caption{Four different fits to the ALMA continuum photometry of HZ10. \textit{Upper left:} our fiducial optically thin fit, also shown in Figure \ref{fig:dataFit}. \textit{Upper right:} a fit with fixed optical depth of unity at $\lambda_\mathrm{thick} = 100\,\mu\mathrm{m}$, which has previously been assumed for HZ10 by \citet{faisst2020} and \citet{villanueva2024}. \textit{Lower left:} a fit with varying $\lambda_\mathrm{thick}$. \textit{Lower right:} an optically thin fit with multi-temperature dust, using the analytical prescription from \citet[][see also Algera et al.\ in preparation]{sommovigo2025}. While the dust temperature depends on the assumption regarding optical depth, all models yield similar dust masses, emissivity indices, and infrared luminosities.}
    \label{fig:dustFitsComparison}
\end{figure*}

As discussed in Section \ref{sec:results}, our fiducial fit to the ALMA photometry of HZ10 is that of an optically thin MBB (Figure \ref{fig:dataFit}). We compare this model to three additional fits in Figure \ref{fig:dustFitsComparison}, adopting either a fixed optical depth of unity at $\lambda_\mathrm{thick} = 100\,\mu\mathrm{m}$ (upper right panel), a varying optical depth (lower left), or optically thin, multi-temperature dust (lower right).

For the fit with fixed optical depth, we adopt $\lambda_\mathrm{thick} = 100\,\mu\mathrm{m}$ following earlier work on HZ10 by \citet{faisst2020} and \citet{villanueva2024}. Compared to our fiducial optically thin fit, we find a larger dust temperature of $T_\mathrm{dust} = 51_{-7}^{+8}\,\mathrm{K}$ (c.f., $T_\mathrm{dust} = 37_{-5}^{+6}\,\mathrm{K}$), consistent with these earlier studies. The inferred dust mass of $\log(M_\mathrm{dust}/M_\odot) = 7.8 \pm 0.2$ is furthermore consistent with our fiducial value of $\log(M_\mathrm{dust}/M_\odot) = 8.0 \pm 0.1$ within the uncertainties. The recovered dust emissivity index of $\beta_\mathrm{IR} = 2.2 \pm 0.3$ is similarly consistent with our fiducial fit.

For the fit with varying optical depth, we adopt a flat prior on the logarithm of the wavelength where the dust turns optically thick of $\log(\lambda_\mathrm{thick}/\mu\mathrm{m}) \in [-1, 3]$. This wide prior is chosen to allow for the possibility of both mostly optically thin dust ($\lambda_\mathrm{thick} \ll 100\,\mu\mathrm{m}$), and an optical depth beyond what was assumed by \citet{faisst2020} and \citet{villanueva2024}. We find that $\lambda_\mathrm{thick}$ cannot be well constrained by our fit, and recover a bimodal distribution with either 1) mostly optically thin dust with a low temperature of $T_\mathrm{dust} \sim 35\,\mathrm{K}$, or 2) dust that is optically thick out to $\lambda_\mathrm{thick}\sim150\,\mu\mathrm{m}$ with a warm temperature of $T_\mathrm{dust} \gtrsim 50\,\mathrm{K}$. The median recovered value of $\log(\lambda_\mathrm{thick}/\mu\mathrm{m}) = 1.44_{-1.65}^{+0.80}$ corresponds to a wide range of possible values of $\lambda_\mathrm{thick} \lesssim 170\,\mu\mathrm{m}$. This translates into large uncertainties on both the dust temperature ($T_\mathrm{dust} = 43_{-9}^{+36}\,\mathrm{K}$) and the dust mass ($\log(M_\mathrm{dust} / M_\odot) = 7.9_{-0.5}^{+0.2}$). However, both of these values are consistent with our fiducial optically thin values.

Finally, we apply a fit with a multi-temperature, optically thin dust distribution following \citet{sommovigo2025}. Briefly, this assumes that the dust temperature distribution of HZ10 can be modeled as a skewed lognormal distribution represented by a mass-weighted dust temperature ($\bar{T}_\mathrm{dust}$), a standard deviation ($\sigma_T$) and skewness ($\xi$). All three parameters, as well as the dust mass and emissivity index, are varied in the fit, adopting the same priors as in Algera et al.\ (in preparation). From this multi-temperature model, we infer a mass-weighted dust temperature of $\langle T_\mathrm{d}\rangle_\mathrm{MW} = 29_{-7}^{+9}\,\mathrm{K}$, and a luminosity-weighted one of $\langle T_\mathrm{d} \rangle_\mathrm{LW} = 40_{-6}^{+9}\,\mathrm{K}$. We moreover infer a dust mass of $\log(M_\mathrm{dust}/M_\odot) = 8.2_{-0.2}^{+0.3}$.

Comparing the four models, we find that the recovered dust temperatures, masses, emissivity indices and infrared luminosities are all consistent with each other within the uncertainties. This motivates our adoption of the optically thin dust model as fiducial, with a dust mass of $\log(M_\mathrm{dust} / M_\odot) = 8.0 \pm 0.1$ -- well within the range obtained from the three other models of $\log(M_\mathrm{dust} / M_\odot) = 7.8 - 8.2$. As such, our analysis of the dust build-up in HZ10 in Section \ref{sec:dustBuildUp} does not strongly depend on which model for its dust emission is used.

\end{appendix}

\end{document}